\documentclass[10pt,letterpaper]{article}
\usepackage[top=0.85in,left=2.75in,footskip=0.75in]{geometry}

\usepackage{amsmath,amssymb}

\usepackage{changepage}

\usepackage{textcomp,marvosym}

\usepackage{cite}

\usepackage{nameref,hyperref}

\usepackage[right]{lineno}

\usepackage[nopatch=eqnum]{microtype}
\DisableLigatures[f]{encoding = *, family = * }

\usepackage[table]{xcolor}

\usepackage{array}

\newcolumntype{+}{!{\vrule width 2pt}}

\newlength\savedwidth

\newcommand\thickhline{\noalign{\global\savedwidth\arrayrulewidth\global\arrayrulewidth 2pt}%
\hline
\noalign{\global\arrayrulewidth\savedwidth}}

\raggedright
\usepackage[aboveskip=1pt,labelfont=bf,labelsep=period,justification=raggedright,singlelinecheck=off]{caption}

\makeatletter
\renewcommand{\@biblabel}[1]{\quad#1.}
\makeatother

\usepackage{lastpage,fancyhdr,graphicx}
\usepackage{epstopdf}
\fancyheadoffset[L]{2.25in}
\fancyfootoffset[L]{2.25in}
\begin{document}
\vspace*{0.2in}

% Title must be 250 characters or less.
\begin{flushleft}
{\Large
\textbf\newline{ Modelling the mitigation of anti-vaccine opinion propagation to suppress epidemic spread: A computational approach  } % Please use "sentence case" for title and headings (capitalize only the first word in a title (or heading), the first word in a subtitle (or subheading), and any proper nouns).
}

% \newline
% Insert author names, affiliations and corresponding author email (do not include titles, positions, or degrees).
% \\
Sarah Alahmadi\textsuperscript{1*}, 
Rebecca Hoyle\textsuperscript{2},
Michael Head\textsuperscript{3}, 
Markus Brede\textsuperscript{1},
% Name5 Surname\textsuperscript{2\ddag },
% Name6 Surname\textsuperscript{2\ddag},
% Name7 Surname\textsuperscript{1,2,3*},
% with the Lorem Ipsum Consortium\textsuperscript{\textpilcrow}
\\
\bigskip
\textbf{1} School of Electronics and Computer Science, University of Southampton, Southampton, United Kingdom
\\
\textbf{2} School of Mathematical Sciences, University of Southampton, Southampton, United Kingdom
\\
\textbf{3} Clinical Informatics Research Unit, Faculty of Medicine, University of Southampton, Southampton, United Kingdom
\\
\bigskip

% Insert additional author notes using the symbols described below. Insert symbol callouts after author names as necessary.
% 
% Remove or comment out the author notes below if they aren't used.
%
% Primary Equal Contribution Note
% \Yinyang These authors contributed equally to this work.

% Additional Equal Contribution Note
% Also use this double-dagger symbol for special authorship notes, such as senior authorship.
% \ddag These authors also contributed equally to this work.

% Current address notes
% \textcurrency Current Address: Dept/Program/Center, Institution Name, City, State, Country % change symbol to "\textcurrency a" if more than one current address note
% \textcurrency b Insert second current address 
% \textcurrency c Insert third current address

% Deceased author note
% \dag Deceased

% Group/Consortium Author Note
% \textpilcrow Membership list can be found in the Acknowledgments section.

% Use the asterisk to denote corresponding authorship and provide email address in note below.
* sha1a21@soton.ac.uk

\end{flushleft}
% Please keep the abstract below 300 words
\section*{Abstract}

Information regarding vaccines from sources such as health services, media, and social networks can significantly shape vaccination decisions. In particular, the dissemination of negative information can contribute to vaccine hesitancy, thereby exacerbating infectious disease outbreaks. This study investigates strategies to mitigate anti-vaccine social contagion through effective counter-campaigns that disseminate positive vaccine information and encourage vaccine uptake, aiming to reduce the size of epidemics. In a coupled agent-based model that consists of opinion and disease diffusion processes, we explore and compare different heuristics to design positive campaigns based on the network structure and local presence of negative vaccine attitudes. We examine two campaigning regimes: a static regime with a fixed set of targets, and a dynamic regime in which targets can be updated over time. We demonstrate that strategic targeting and engagement with the dynamics of anti-vaccine influence diffusion in the network can effectively mitigate the spread of anti-vaccine sentiment, thereby reducing the epidemic size. However, the effectiveness of the campaigns differs across different targeting strategies and is impacted by a range of factors. We find that the primary advantage of static campaigns lies in their capacity to act as an obstacle, preventing the clustering of emerging anti-vaccine communities, thereby resulting in smaller and unconnected anti-vaccine groups.  On the other hand, dynamic campaigns reach a broader segment of the population and adapt to the evolution of anti-vaccine diffusion, not only protecting susceptible agents from negative influence but also fostering positive propagation within negative regions.

% Please keep the Author Summary between 150 and 200 words
% Use first person. PLOS ONE authors please skip this step. 
% Author Summary not valid for PLOS ONE submissions.   
% \section*{Author summary}
% Lorem ipsum dolor sit amet, consectetur adipiscing elit. Curabitur eget porta erat. Morbi consectetur est vel gravida pretium. Suspendisse ut dui eu ante cursus gravida non sed sem. Nullam sapien tellus, commodo id velit id, eleifend volutpat quam. Phasellus mauris velit, dapibus finibus elementum vel, pulvinar non tellus. Nunc pellentesque pretium diam, quis maximus dolor faucibus id. Nunc convallis sodales ante, ut ullamcorper est egestas vitae. Nam sit amet enim ultrices, ultrices elit pulvinar, volutpat risus.

\nolinenumbers

% Use "Eq" instead of "Equation" for equation citations.
\section*{Introduction}
Throughout history, diseases have posed a constant threat to human health and wellbeing, with the COVID-19 pandemic serving as a recent and striking example. Vaccination is a vital tool for combating disease prevalence. However, in many instances, the availability of vaccinations does not necessarily lead to their uptake owing to several factors, including vaccine hesitancy. According to the WHO, vaccination hesitancy is one of the top ten threats to public health  \cite{WHO2019}. One significant factor contributing to it is exposure to vaccine misinformation \cite{dube2013vaccine}. As a result, such misinformation may spread between people, leading to negative collective behaviour towards vaccination due to peer influence \cite{zhang2017preferential}. Therefore, achieving a high uptake of vaccination 
represents a social dilemma for public health administration \cite{zhang2010hub, zhang2014effects, ichinose2017positive, wang2020vaccination}. The analysis and investigation of the dilemma of vaccination has been studied extensively from various perspectives. For example, vaccination decision behavior has been studied with respect to vaccine-related information diffusion \cite{salathe2008effect,funk2009spread,ruan2012epidemic,campbell2013complex,da2019epidemic,mehta2020modelling,yin2022impact},  with respect to rationality in social imitation using evolutionary game theory \cite{fu2011imitation,cardillo2013evolutionary,zhang2014effects,ichinose2017positive,zhang2017preferential}, and with respect to a combination of both approaches \cite{meng2022analysis,yin2022impact}. 

Information about a disease and attitudes around vaccination play a significant role in people's willingness to get vaccinated and, therefore, the extent and severity of an outbreak \cite{dube2015vaccine}. COVID-19 is the most recent example of the influence of misinformation on epidemic dynamics  \cite{sontag2022misinformation}. Studying misinformation and vaccine opinion diffusion not only helps in understanding the range of influences on people's attitudes, but also helps in understanding the flow of attitudes across the social network and, therefore,  allows for more effective intervention control strategies. Many studies have investigated the role of information dissemination on the scale of transmission of an epidemic. This information may positively serve as a trigger for self-protection, for example when awareness of a disease spreads, as in \cite{ruan2012epidemic,funk2009spread,da2019epidemic,sontag2022misinformation}, where the authors investigate how the spread of awareness may reduce the size of an epidemic. On the other hand, it may work as a stimulant for negative behavioural responses, such as the spread of false information regarding vaccinations over a social network, which negatively influences people's inclination to vaccinate \cite{salathe2008effect, campbell2013complex,dorso2017vaccination,dai2019coevolution,mehta2020modelling,mumtaz2022exploring,chen2022coevolving,yin2022impact}. 

Opinions regarding vaccines spread among individuals as a social contagion and significantly affect the distribution of vaccination coverage, subsequently influencing the transmission dynamics of diseases. Previous studies have explored the interplay between disease spread and the propagation of vaccine-related opinions. Some studies have focused on dynamics of anti-vaccine opinions while neglecting the influence of pro-vaccine propagation \cite{campbell2013complex, mumtaz2022exploring, chen2022coevolving}. Others have examined both anti-vaccine and pro-vaccine dynamics, with anti-vaccine dynamics being treated as a contagion phenomenon \cite{mehta2020modelling, dorso2017vaccination}, and a limited number of studies considered the propagation of both anti- and pro-vaccine behaviour \cite{salathe2008effect, dai2019coevolution}. In addition, researchers have employed a range of techniques to model opinion transmission. For instance, epidemic models have been used in \cite{mehta2020modelling, mumtaz2022exploring, chen2022coevolving, yin2022impact}, while others have employed opinion diffusion models, such as threshold-based model \cite{campbell2013complex}, voter-like model \cite{salathe2008effect}, the m-model \cite{dai2019coevolution}, and a variant of Axelrod’s model \cite{dorso2017vaccination}. The utilization of epidemiological models to describe the spread of opinions provide simple contagion, where an individual can be influenced by a single exposure. In contrast, empirical studies have demonstrated that influence propagation exhibits complex contagion diffusion pattern \cite{centola2010spread,romero2011differences}, where multiple exposures are required to influence an individual.  Consequently, opinion diffusion models are better suited to capture the dynamics of social interactions, as they incorporate  more  realistic behavioral characteristics that govern the adoption of opinions and decision-making in the real world.
  
Anti-vaccine opinion adopters tend to cluster in social networks \cite{salathe2011assessing,dube2015vaccine,omer2009vaccine}. This leads to the existence of clusters of unprotected individuals which represent a threat to public health and prevent attainment of high vaccination
rates \cite{lieu2015geographic,yuan2019examining} as well as impeding the effects of full herd immunity and high vaccination rates \cite{salathe2008effect,gromis2022spatial}. Numerous studies have investigated the risk of anti-vaccine clusters, focusing for instance on, the relationship between anti-vaccine communities and the epidemic size \cite{salathe2008effect,salathe2011assessing,campbell2013complex,dorso2017vaccination},  understanding the communication between vaccine social clusters \cite{yuan2019examining}, correlation between spatial clusters and the associated vaccination rates \cite{lieu2015geographic}, and using social media data to identify vaccine opinion clusters and vaccination rates  \cite{bello2017detecting,salathe2011assessing,cheng2023exploring}.

On the other hand, regarding only the spread of information, mitigating the propagation of misinformation has received  researchers' attention in the field of influence minimization, which is a subclass of the influence maximization problem. The influence minimization problem focuses on minimizing the propagation of undesirable influence in a
social network. Researchers have addressed this problem  either by blocking specific nodes \cite{pham2020multi} or edges \cite{zareie2022rumour} in the network, or  applying true information campaigns \cite{budak2011limiting,liu2016containment,tong2018misinformation,yang2019rumor,yang2020containment,zhang2015limiting}. In the context of vaccination hesitancy, empirical research highlights the significance of healthcare providers' advice in  overcoming vaccination hesitancy and promoting vaccine uptake through the dissemination of  accurate information about vaccines  \cite{adeyanju2021determinants,marzo2022perceived}.  In addition, it is worth noting that the main focus of existing research on misinformation mitigation has primarily centered around reducing the number of negative opinion adopters or increasing the number of positive opinion adopters. In the present study, our focus encompasses not only the reduction of anti-vaccine adopters but also their distribution in the network, aiming to mitigate the growth and connectivity of anti-vaccine communities.

Although there is existing literature on the dynamics of vaccine-related information and opinion diffusion and their impact on epidemic spread \cite{salathe2008effect,campbell2013complex,dorso2017vaccination,dai2019coevolution,yin2022impact}, several gaps have been identified in the field. First,  there is limited consideration of both pro- and anti-vaccine social interactions and preferences that drive epidemic dynamics. Second,  a research gap exists on the mitigation of anti-vaccine contagion and its implications for the size of epidemics. Additionally, despite many studies emphasizing the risk posed by anti-vaccine communities and their correlation with the extent of disease spread, a closer examination of the literature reveals the need for research focused on mitigating the expansion and connectivity of anti-vaccine clusters. Further investigation in these areas could be valuable for advancing our understanding and enhancing our ability to manage and control the spread of diseases.

Owing to the correlation between the epidemic spread dynamics and the vaccine behaviour diffusion, we develop a computational framework to examine the effects of mitigating anti-vaccine opinions on the size of epidemics. It involves implementation of a strategic counter-campaign that effectively disseminates positive information about vaccines to counteract the diffusion of anti-vaccine opinions. This model integrates vaccine-related information dissemination, pro- and anti-vaccine  opinion diffusion, and disease spread. Our main contributions are as follows: 
\begin{enumerate}
    \item  We address the mitigation of the diffusion of anti-vaccine opinions as a misinformation mitigation problem, where the primary objective is to seek an optimal set of targets to seed positive influence in a social network. To the best of our knowledge, the misinformation mitigation problem has not been studied in the context of coupled models, particularly in the field of coupled dynamics of vaccine-related information and disease spread. Thus, we explore and compare different heuristics seeking to identify a set of nodes within a network and investigate their effects on the structure of anti-vaccine communities and, ultimately, the extent of disease spread.
    
    \item We consider two approaches for the seed set selection: static and dynamic. In the static approach, which  has traditionally been considered \cite{budak2011limiting,yang2019rumor}, the seed set is chosen at the beginning of the campaign launch and remains unchanged throughout the campaign. In contrast, in the dynamic approach, the seed set is selected repeatedly in different rounds based on specific criteria such as \cite{wijayanto2019effective,shi2019adaptive}. We propose a novel dynamic approach where we update the target set with a new target set based on certain criteria. The selection criteria in this approach are based on the local presence of negative vaccine-related information. This is an adaptive approach that responds to the evolving dynamics of the anti-vaccine propagation.

\end{enumerate}

 Experiments reveal that different methods of  distributing positive influence have varying impact on anti-vaccine opinion diffusion and  the structure of anti-vaccine communities which consequently affects the epidemic spread. For instance, targeted campaigns demonstrate higher efficiency compared to random campaigns, and the dynamic approach is more efficient than the static approach. Nonetheless, depending on other factors: the allocated positive budget, the level of social influence between individuals, the size of the target set, and the time horizons of dynamic campaigns, we observe variations in the effectiveness of controlling  anti-vaccine propagation for each strategy. In the results section, we provide a comprehensive analysis of the performance of each campaign.

\section*{Model description and methods}

In this study, we consider a scenario where opinion exchanges and the vaccination process occur prior to the spread of the disease. This type of modelling is particularly relevant for diseases like childhood illnesses, e.g., measles, where vaccinations typically take place during the early years of childhood, and the disease manifests during school or preschool age. As a relevant real-world example, in 2019, after almost two decades of significant progress in global vaccination programs, there was a resurgence of measles \cite{hotez2020combating}. One cause of this resurgence is due to a significant decline in vaccine coverage, resulting from the spread of anti-vaccine information. 

 Motivated by the model presented in \cite{campbell2013complex}, we developed a framework consisting of a two-stage agent-based model. The first stage involves  opinion diffusion and vaccination processes in relation to individuals' opinions about vaccines, and the second stage is the disease spread among unvaccinated individuals. This modeling approach is a common methodology found in the existing literature on vaccination programs for childhood diseases \cite{salathe2008effect} and flu-like diseases \cite{ichinose2017positive, zhang2017preferential, fu2011imitation,tatsukawa2021free}. We assume that the vaccination provides full immunity, implying that all who receive it are immune to the disease. For vaccine opinion diffusion, motivated by the approach of \cite{campbell2013complex}, we developed a model for the dual propagation of positive and negative opinions. Similarly, as in \cite{campbell2013complex}, we utilized the SIR model for the disease spread stage, originally developed by \cite{kermack1927contribution}, which is widely used by researchers in epidemic modeling.
 
The experiment workflow commences with the stage of opinion diffusion, during which sentiments related to vaccines spread in the network, subsequently influencing individuals' vaccination decision-making. As influenced individuals adopt particular opinions, their social influence starts to propagate through the network. This social contagion inevitably leads to the formation of homogeneous communities, i.e. a community with a particular attitude towards vaccination, with anti-vaccine communities being our primary concern. In our model, we utilize complex contagion, which implies that agents need multiple exposures to be influenced, as this phenomenon has also been observed in influence propagation processes \cite{centola2010spread,romero2011differences}. Following the stage of opinion dissemination, the vaccination process takes place. Subsequently, the infectious disease begins to spread among unimmunized individuals. The details of each stage will be explained in the following sections.

\begin{figure}[!h]

	\begin{minipage}[t]{\linewidth}
	\centering         
	\includegraphics[scale=0.4]{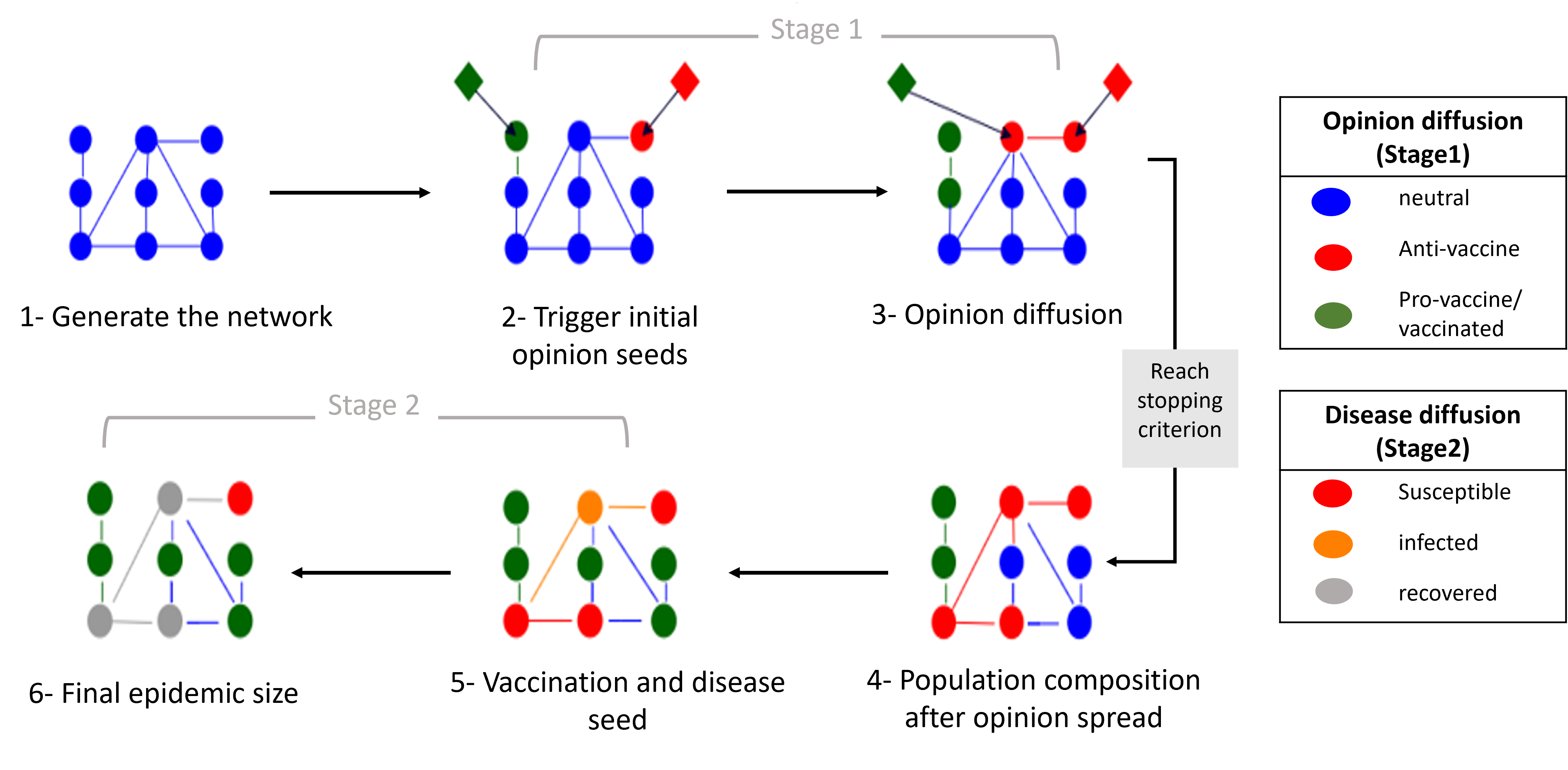}

	\end{minipage}

\captionsetup{justification=justified, singlelinecheck=false}
\caption{{\bf Illustration of the model describing opinion formation and disease propagation.} Blue circles represent neutral individuals, green circles represent pro-vaccine individuals, and red circles represent anti-vaccine individuals. The first stage involves the generation of the social network and the initialization of agent opinion states as agents with neutral opinions. Then, external exposures to positive and negative information triggers the initial seed sets for both anti-vaccine and pro-vaccine contagion. Opinion diffusion continues  until a stopping criterion is reached. In this stage, a vaccination takes place for all non-negative individuals. Subsequently, a randomly chosen non-vaccinated individual is infected, and the spread of the disease continues until no further newly infected agents are generated. Finally, we record the number of recovered agents to measure the epidemic size.}

\label{Fig1} 
\end{figure}

We conduct our experiments using the Watts-Strogatz small-world network model \cite{watts1998collective}. Small-world network provides an effective framework for modelling complex systems, as many real-world networks including social networks \cite{barrat2000properties}, exhibit the small-world property \cite{latora2001efficient}. In addition, as  the model is stochastic in nature, we conduct a number of simulations per scenario to obtain the average epidemic size. In each simulation,  we generate the contact network, followed by the diffusion of opinions and vaccination, and finally the disease spread. An illustration of the model stages is shown in Fig \ref{Fig1}.

\subsection*{Opinion diffusion} \label{opin-stage}
Our model consists of a network G  composed of N nodes,  represented by $G(V,E)$, where $V$ is the set of nodes representing individuals, $V= \{1,2,...N\}$, and  $E$ is the set of edges representing contacts between individuals. We assume that this contact structure is the same for both the flow of information and the transmission of the disease. We consider two types of vaccine-related exposures: positive, which spread positive sentiment, and negative, which spread negative sentiment. In addition, we consider exposures from external sources, referred to as general exposure or campaigns, occurring with probabilities $\mu^-$ (negative) and $ \mu^+$ (positive), as well as through social communication where influence is exerted by opinion adopters on each of their neighbours with probabilities $\omega^-$ (negative) and $ \omega^+$ (positive), per timestep. Each agent has their own set of counters $\{ \phi^-, \phi^+ \}$, where  $\phi^-_i$ quantifies exposures to negative, and $\phi^+_i$ quantifies exposures to positive sentiments experienced by each agent.  Furthermore, $\theta$ is an opinion decision threshold that represents an individual's sensitivity to influence. We assume that an agent shifts its opinion from a neutral state to either negative or positive when it has been exposed to $\theta$ more exposures of a particular influence. 

Each agent $i$, $i=\{1,2,..,N\}$, may adopt one of three opinion states  $ s_i \in \{o^-, o^0, o^+ \} $, where  $o^-$ is negative, $o^+$ is positive, and  $o^0$ is neutral. We assume that once an agent changes its state from a neutral to a negative (or positive) state, it remains in that state. To summarize, following a setup informed by the approach taken in  \cite{campbell2013complex}, our model operates in discrete time steps as follows:
\begin{itemize}
 \item At time $t=0$, all agents are neutral, i.e., ${   s_i = o^0 , \forall i \in V }$. 
 
 \item At each time step $t$, negative general exposure exerts influence on the population with probability $\mu^-$ per individual. Similarly, positive general exposure exerts influence on the targeted population with probability $\mu^+$ per individual.  In addition, each agent $i$ with state $s_i \in \{o^-, o^+ \}$, exerts influence on each of its neutral neighbors with probability $\omega^{-}$ for negative opinion adopters and $\omega^{+}$ for positive opinion adopters. Exposures for an agent $i$, i.e, $\{i \in V|s_i=o^0\}$,  is added to $\phi^{-}_i$ if it is negative or added to $\phi^{+}_i$ if it is positive.
 
 \item At each time step $t$, each neutral agent $i$, i.e, $\{i \in V|s_i=o^0\}$, updates its opinion state as follows: \\ 
    \begin{equation}
    \label{opinion decision}
    \begin{aligned}
   s_i= \begin{cases}
    o^- &\text{if $\phi^-_i - \phi^+_i \geq \theta$}, \\
    o^+ &\text{if $\phi^-_i - \phi^+_i  \leq  - \theta$},\\
    o^0 &\text{otherwise.}\\
    
    \end{cases}
    \end{aligned}
\end{equation}

 \end{itemize}

\begin{figure}[!h]

	% \begin{minipage}[t]{0.8\linewidth}
	\centering         
	\includegraphics[scale=0.4]{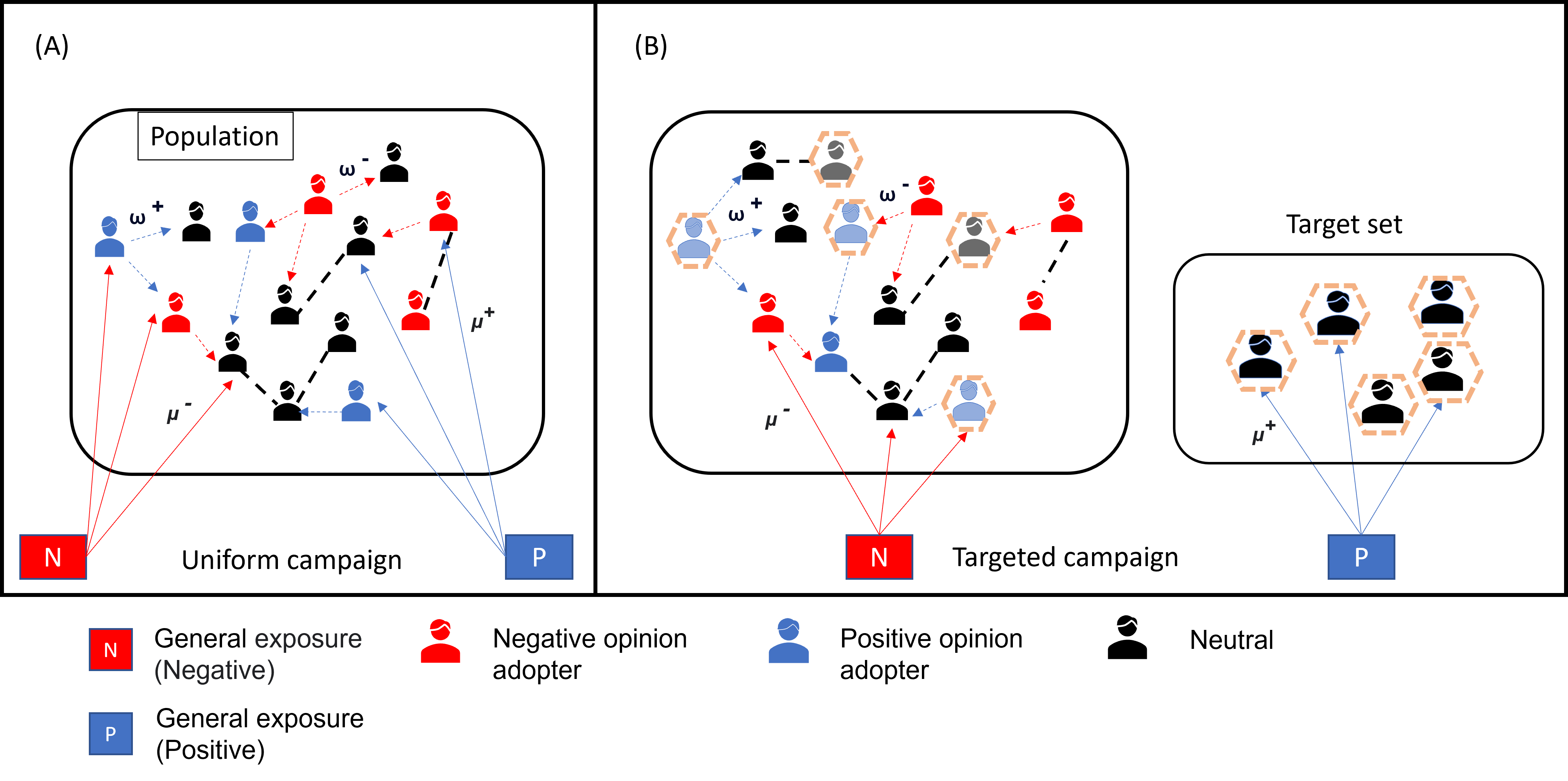}

	% \end{minipage}

\captionsetup{justification=justified, singlelinecheck=false}
\caption{{\bf Illustration of opinion propagation and campaigning methods.} The figure shows the exchange of vaccine-related opinions and external exposures, as well as the positive campaign types. (A) Random dissemination of negative and positive vaccine-related sentiments from  external campaigns to the public. (B)  Targeted positive campaign. $\mu^-$, and $\mu^+$ are the general exposure rates for negative and positive sentiments,
respectively. $\omega^-$, $\omega^+$ are the social exposure rates for negative and positive opinions,
respectively.}
\label{Fig2} 
\end{figure}

The above process is repeated for $\tau$ steps. A visual depiction of external campaigns and social influence is shown in Fig \ref{Fig2}A. Next, we make the assumption that all agents with a non-negative opinion will receive the vaccine, while those with a negative opinion will refuse it.

\subsection*{Epidemic spread} \label{SIR-stage}
In this phase, an anti-vaccine opinion adopter is randomly chosen as a seed for the disease, and disease spread is modeled using the SIR model. Each infected agent can transmit the disease with a probability $\beta$ per contact per time step, and can recover with a probability $\gamma$ per time step. The process continues until there are no more infected individuals, and the epidemic size $S_r$ is recorded.  $S_r$ is defined as the total number of individuals who experience an infection during the course of the epidemic \cite{house2013big}.

Table ~\ref{tableParm} below presents a list of the model parameters and their corresponding descriptions.
\begin{table}[!ht]
% \begin{adjustwidth}{-2.25in}{0in} % Comment out/remove adjustwidth environment if table fits in text column.
\centering
\caption{
{\bf Model parameters and descriptions.} 
% This table presents a list of the model parameters that govern propagation of the opinion diffusion and disease spread phases, along with a brief description of each parameter.
}
\begin{tabular}{|p{0.25\textwidth} |p{0.7\textwidth}|} 
\hline
{\bf Parameter} & {\bf Parameter Description}\\ \thickhline
$N$ & Population size \\  \hline
$\mu^-$ & Negative general exposure rate \\ \hline
$\mu^+$ & Positive general exposure rate \\ \hline
$\omega^-$ & Negative social rate \\ \hline
$\omega^+$ & Positive social rate \\ \hline
$\phi^-$ & Negative exposure counter \\ \hline
$\phi^+$ & Positive exposure counter \\ \hline
$\theta$ & Opinion formation threshold \\ \hline
$T$ & Target set size \\ \hline
$t_r$ & Update time interval for the dynamic control \\ \hline
$\tau$ & Maximum number of time steps \\ \hline
$\beta$ & Disease infection rate \\ \hline
$\gamma$ & Disease recovery rate \\ \hline
$I_0$ & Initial number of infected individuals \\ \hline
$S_r$ & Epidemic size \\ \hline

\end{tabular}
% \begin{flushleft} 
% \end{flushleft}
\label{tableParm}
% \end{adjustwidth}
\end{table}

\section*{Campaign strategies} \label{pos-camps}
In this study, we define a positive campaign as a strategic allocation of the strength of positive external vaccine information, denoted as $\mu^+_i$,   to the agents $i=1,...,N$, with $1/N\sum_i^N \mu^+_i = \mu^+$.  We compare the effectiveness of various types of such positive campaigns against a random negative campaign that spreads negative vaccine information, assuming that each agent can be negatively influenced with an influence strength $\mu^-_i=\mu^-$ at each time step. This section outlines the proposed selection strategies for the target set for the positive campaign.

\subsection*{Random campaign (UnifRand)} \label{randStrgy} 
In this campaign, all individuals are exposed to the general positive exposure with a uniform positive allocation $\mu^+_i = \mu^+ $ at each time step. In this scheme, we extend the work introduced by \cite{campbell2013complex}, in which they explored the impact of only a negative campaign and the diffusion of anti-vaccine opinions on the configuration of anti-vaccine communities. The campaign scheme is illustrated in Fig \ref{Fig2}(A).

\subsection*{Targeted campaigns} 
In this approach, we aim to target a certain set of agents with neutral opinions, as illustrated in Fig \ref{Fig2}(B), to efficiently mitigate the spread of anti-vaccine influence. This puts a greater emphasis on specific individuals and has been demonstrated as an effective method, as evidenced by empirical studies \cite{konstantinou2021transmission}. There are several ways to target them: (\romannumeral 1) random selection as a common and intuitive approach; (\romannumeral 2) based on their topological position on the network; (\romannumeral 3) based on their neighbourhood status with regards to local information about vaccine opinions. Each of these will be explained in detail in the subsequent  sections.  

Furthermore, this approach involves two types of campaigning: static and dynamic. In the static approach, the target set is selected based on predetermined criteria prior to the launch of the campaign and this set remains unchanged. In the dynamic approach, the initial targets are selected at random, however, every $t_r$ opinion updates the target set is updated and replaced by new targets. 

In the targeted approach, let $T_i$ =1 if agent $i$ is targeted and $T_i =0$ otherwise. Accordingly,  the positive campaign allocation  $\mu^+_i =T_i \times \mu^+ \times N / \sum_i^N T_i$.
In the following, we have explored and compared the following six heuristics for selecting the target set. 

\subsubsection*{Static campaigns} 
\label{statc-camps}
The following strategies outline the criteria for selecting the target set in the static campaign approach.
\begin{enumerate} %[label=(\roman*)]
%Erdős–Rényi    
\item \textbf{Targeted random strategy (TargtRand)}: a random subset of the entire population is selected as the campaign's targets.
\item\textbf{Targeted centrality-based strategy (Cntrl)}: this is a topology-based campaign in which targets are selected based on betweenness centrality. The $T$ most central agents in the network are targeted. In the case of ties among agents with the same score, we randomly select T agents from among the tied nodes.  We compute the betweenness centrality score for each agent following the algorithm proposed by \cite{brandes2001faster}. Betweenness centrality measures the extent to which a node lies on the shortest paths between other nodes in the network \cite{latora_nicosia_russo_2017}. Nodes with high betweenness centrality are considered to be mediators \cite{zhang2017degree} and are often located on important bridges in the network, making them key players for information flow and communication. Targeting these nodes can potentially have a greater impact on the overall network dynamics, creating barriers for anti-vaccine communities and preventing them from merging together.
\end{enumerate}   
\subsubsection*{Dynamic campaigns} \label{dync-camps}
% \begin{description} [leftmargin=5pt]
Dynamic campaign strategies rely on the local information about individuals' opinions regarding vaccines. The following strategies outline the criteria for selecting the target set based on this information.
\begin{enumerate} %[label=(\roman*)]   
\item\textbf{Dynamic random strategy (DynRand)}:  at each time interval \emph{$t_r$}, the target set is replaced with a new target set by selecting from the remaining population of agents with neutral opinions at random. 
We consider this campaigning strategy as a reference to evaluate the effectiveness of other dynamic selection criteria.
\item\textbf{Dynamic local information based strategies}: in this approach, our objective is to focus on agents with neutral opinions who are susceptible to negative influence from their social connections, i.e., agents with neutral opinion who have at least one anti-vaccine neighbour. This is a neighborhood-based scheme with two primary considerations: first, placing seeds to effectively inhibit the growth of the negative cluster, which requires them to have a certain number of negative neighbors; second, situating  positive seeds in a way that maximizes the potential for positive clusters to grow (and eventually block negative clusters), which means targeting neutral agents with the greatest number of neutral neighbors. We expect a trade-off here. If agents have too many negative neighbors, they may become overwhelmed quickly, and thus positive influence might be wasted. Contrariwise, targeting agents with too many neutral neighbors might place them too far from negative clusters, thus becoming inefficient at blocking negative clusters from growing. We formalise this trade-off  as follows:  

\begin{enumerate} 
\item\textbf{ Local information based (Locl-Info)}: 
at each time interval $t_r$, the target set is updated and replaced with a new target set by selecting at random from the remaining population of agents with neutral opinions who have at least one anti-vaccine neighbor.  

\item \textbf {Advanced Locl-Info with single-objective (advLocl-Info):} Here, we aim to target neutral agents who are in neighbourhoods that meet a trade-off between blocking negative influence and allowing positive influence to spread. The first is related to the number of adjacent negative agents and the second relates to the number of adjacent neutral agents. Let $\zeta$ be the target number of anti-vaccine neighbours, here, we seek to target neutral agents who have as close as possible to $\zeta$. In more detail, we do this by scoring agents according to the difference in their number of anti-vaccine neighbours from $\zeta$ as follows: \\ 
    \begin{equation}
    \label{closness equation}
    \begin{aligned}
       {g_i(\zeta)= \lvert n_i^- - \zeta \rvert},
    \end{aligned}
\end{equation} 

where $n_i^-$ denotes the actual count of anti-vaccine neighbours for an agent i. Then, we select the $T$ agents with the lowest score (and selecting at random in case of ties). If the number of agents selected  is less than T, the selection process continues by choosing from the remaining population of agents with neutral opinions at random until T agents have been selected. We will vary $\zeta$ to identify the heuristic that best suppresses disease outbreaks.\label{advLocl-Info}

\item\textbf{Advanced local-info with multi-objective (advMultLocl-Info)}: This heuristic builds on the previous heuristic, but we now also include the potential for positive information to spread by including the number of neutral neighbours in the scoring process. Again, let $\zeta$ be the target number of anti-vaccine neighbours and $Z$ be the target number of neutral neighbours of an agent. Next, presuming that agent $i$ has $n^-_i$ anti-vaccine neighbours and $n^0_i$ neutral neighbours, we calculate a score according to:
\begin{equation}
 \label{multi-obj closness equation}
    \begin{aligned}
       g_i(\zeta,Z) = \lvert  n^-_i - \zeta\rvert +\lvert n^0_i - Z\rvert,
    \end{aligned}
\end{equation}
and select the $T$ agents with the lowest scores. In the case of ties among agents with the same score, we randomly select T agents from among the tied nodes. If the number of agents selected  is less than T, the selection process continues by choosing from the remaining population of agents with neutral opinions at random until T agents have been selected. Below we will explore the dependence of the effectiveness of the heuristic on both target numbers of negative neighbours $\zeta$ and neutral neighbours $Z$. \label{advMultLocl-Info}

 \end{enumerate}
\end{enumerate}

\section*{Results} \label{results-pos-camps}
In this section, we present the obtained epidemic size for the proposed positive campaigns. During the first stage, i.e., opinion exchanges, we consider two cases as stopping criteria: one where opinions spread until all agents have adopted an opinion, referred to as $\tau=\infty$, which represents the long-run scenario and enables the evaluation of the long-term behavior. The other scenario where opinions spread over a certain period of time $\tau$, represents the short-run scenario. In the short-run case, we compared the obtained epidemic size with that from a scenario where only anti-vaccine opinion diffusion is considered, previously investigated in\cite{campbell2013complex}. While our experimental setting differs from that of \cite{campbell2013complex}, we have applied their work within our setting for comparison purposes. We conducted extensive experiments to investigate the factors that determine the efficacy of each campaign in promoting vaccination.

Unless otherwise stated, the results show the epidemic size  as a function of the social contagion rate parameter $\omega$ to compare the varying strengths of social influence on vaccine decision-making and their impact on the emergence of anti-vaccine communities and consequently disease spread.  The social rate is a crucial component because it controls the growth of anti- and pro-vaccine communities. A low social rate implies that individuals are barely influenced by their social contacts, resulting in a low growth of anti-vaccine communities. In contrast, a high social rate indicates that individuals are highly influenced by their social contacts, leading to a large growth of anti-vaccine communities.  To simplify the analysis, we have assumed equal social rates for pro- and anti-vaccine diffusion and evaluated the impact of different rates of  positive general exposure $\mu^+$ on the epidemic size, while keeping the negative general exposure $\mu^-$ fixed. The shaded area and error bars represents the 95$\%$ confidence intervals.

In all experiments, following a similar configuration used in \cite{campbell2013complex} we consider a small-world network with size $N=5000$, rewiring probability $p=0.01$, and average degree $\langle k\rangle=10$. We use an opinion formation threshold of $\theta$ =2. For SIR parameters we use infection rate $\beta =0.1$,  recovery rate $\gamma =0.1$, and seed set $I_0 =1$. Unless otherwise stated, for each scenario we generate 500 different networks and for each network we run 500 SIR infection simulations.

\subsection*{Results of the random campaign} 
\begin{figure}[!h]

% \begin{minipage}[t]{\linewidth}
      % \raggedleft  
      \centering
	\includegraphics[width=\linewidth]{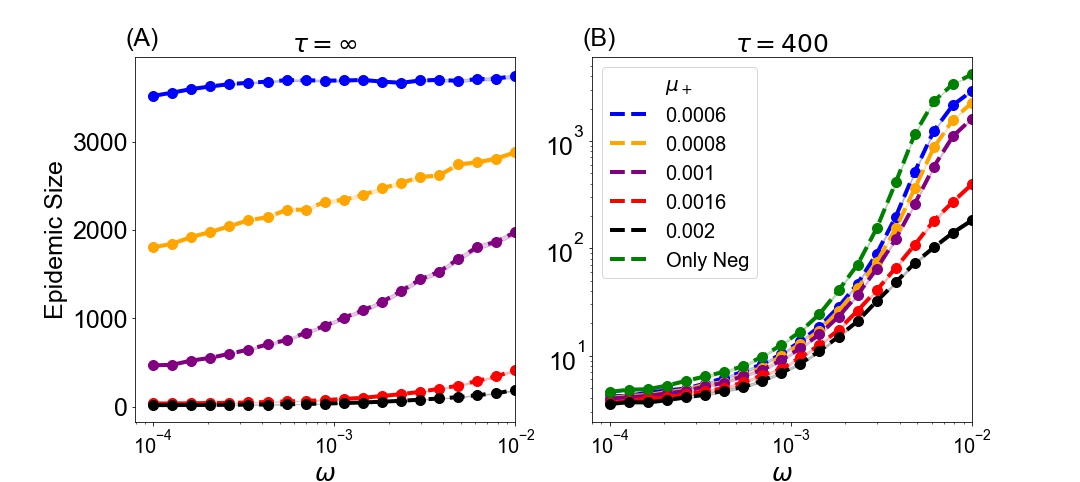}

	% \end{minipage}

\captionsetup{justification=justified, singlelinecheck=false}
\caption {{\bf Average epidemic size for the random campaign as a function of the social rate $\omega=\omega^+=\omega^-$.} (A) $\tau=\infty$ (B) $\tau=400$. The figure shows results for different positive exposure rates $\mu^+$ with fixed negative exposure rate $\mu^-$ = 0.001.}
\label{Fig3}
\end{figure}

This section presents the epidemic size obtained by applying the random campaign to disseminate positive vaccine-related information, as illustrated in Fig \ref{Fig3}.  

Fig \ref{Fig3}A illustrates the long-run scenario with $\tau=\infty$ for varying social rates. This campaign seeds positive influence widely by targeting the whole population at random, resulting in a large number of seeds being generated, with the same occurring for negative seeds. When the social rate $\omega$ is low, the social influence of these seeds is slow, resulting in unconnected and smaller homogeneous communities, namely anti-vaccine and pro-vaccine communities. In addition,  the existence of pro-vaccine communities prevents the merging of anti-vaccine communities.  This ultimately leads to a smaller epidemic size compared to a higher social rate.

In contrast, when the social rate is high, the social influence of these seeds spreads rapidly, leading to fast expansion of the communities and allowing merging of communities. In this scenario, due to the rapid diffusion of opinions, the positive campaign fails to exert more influence over time, as the majority of individuals have already adopted an opinion. This ultimately results in large-sized communities and subsequently a higher epidemic size within anti-vaccine communities. This pattern is observable when the negative and positive general exposure rates diffuse at nearly equal rates, represented by yellow and purple lines in the figure. Nevertheless, with a much lower positive general exposure rate $\mu^+ \ll \mu^-$ (blue line), this pattern is almost nonexistent, since the positive seeding rate is low, rendering the negative influence dominant regardless of the social rate. In contrast, when the positive rate is much greater than the negative rate $\mu^+ \gg \mu^-$, represented by red and black lines, the epidemic size dramatically decreases to less than $50$ at the lowest social rate, i.e., $\omega= 10^{-4}$. Although the social rate does not play a significant role in such scenarios, higher social rates, namely $\omega= 10^{-2}$, lead to an increase in the epidemic size to less than 500.

Moreover, an increase in the positive external rate $\mu^+$ results in a decrease in the epidemic size. The disparity in budget allocations for negative and positive external rates plays a pivotal role in shaping the prevalence of anti-vaccine opinions. A higher positive external rate than negative external rate leads to a dominance of positive influence, resulting in a smaller number of anti-vaccine opinions and consequently a smaller epidemic size, and vice versa.

Fig \ref{Fig3}B gives results for $\tau=400$, and allows for a comparison between scenarios with and without positive campaigns. The figure distinctly illustrates that the propagation of positive vaccine-related information always yields a positive effect, resulting in a reduction in  epidemic size compared to the scenario in which only anti-vaccine opinions are being spread (green line). Furthermore, the suppression of the epidemic increases as the positive general exposure rate $\mu^+$ increases. However, as the social rate increases, the growth of communities also increases, leading to a higher epidemic size. 

\subsection*{Results of the static campaigns}
This section presents the epidemic size obtained by using targeted static campaigns to disseminate positive vaccine-related information. This includes the targeted random and targeted centrality-based campaigns. 

\begin{figure}[!h]

    \centering
    \includegraphics[width=\linewidth]{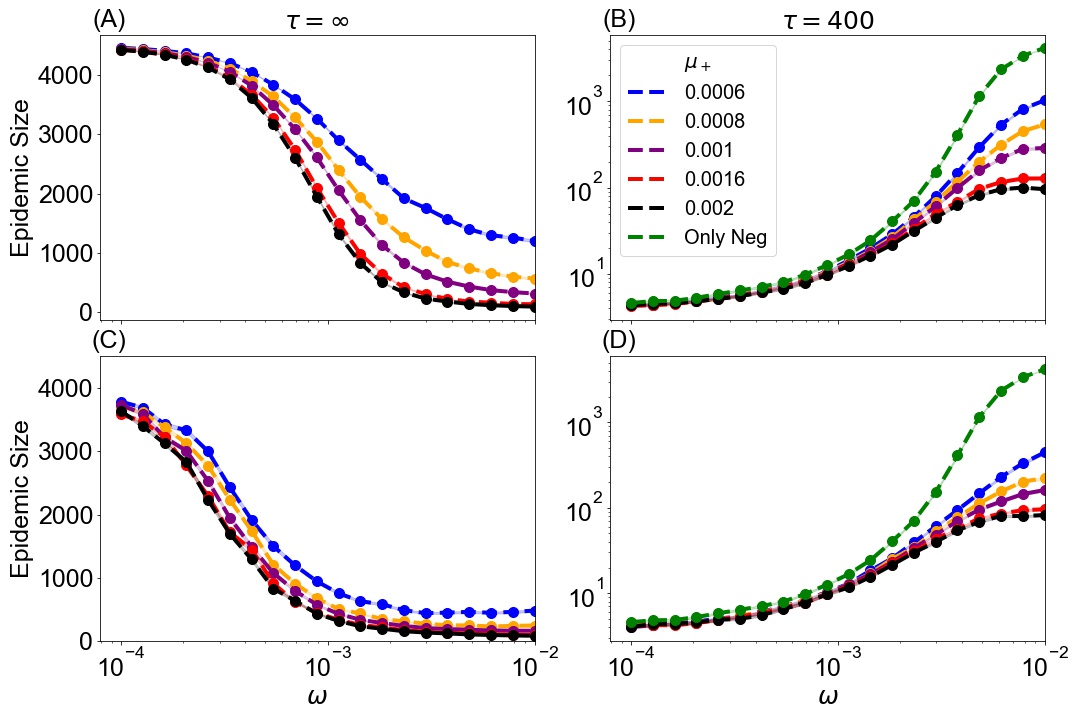}

\captionsetup{justification=justified, singlelinecheck=false}
\caption { {\bf  Average epidemic size for static campaigns as a function of the social rate $\omega=\omega^-=\omega^+$.} (A) and (B) for the targeted random campaign. (C) and (D) for the centrality-based campaign. (A) and (C) $\tau=\infty$,  (B) and (D)  $\tau=400$.  The figures show different positive exposure rates $\mu^+$ with fixed negative exposure rate $\mu^-$ = 0.001. Target set size is $T =500$.}
\label{Fig4}
\end{figure}

Fig \ref{Fig4}A and Fig \ref{Fig4}C illustrate the long-run scenario with $\tau = \infty$ for the targeted random and centrality-based campaigns, respectively. In general, the centrality-based campaign  performs better in reducing the epidemic size than the targeted random campaign. Additionally, as the positive rate $\mu^+$ increases, the performance of the campaigns improves, leading to higher epidemic suppression. However, for the centrality-based campaign, this improvement is relatively small due to the fact that this campaign is efficient even when positive exposure rates are low.  

More importantly, there is a notable difference between the random and targeted campaigns in regard to the social influence rate $\omega$. When the social rate is low, individuals tend to exchange opinions less frequently, and their vaccination behavior is largely influenced by the external campaigns. This is shown clearly in the long-run setting, where all individuals adopt a vaccine opinion, as seen in Figs \ref{Fig3}A,  \ref{Fig4}A, and \ref{Fig4}C. Under these circumstances, the random campaign, see Fig \ref{Fig3}A, tends to yield a smaller  epidemic size compared to static campaigns, see Figs \ref{Fig4}A, \ref{Fig4}C. This is due to the fact that the random campaign generates a larger number of positive seeds over time than the targeted campaigns, which are restricted to a fixed set of agents. With lower levels of social interaction, the growth of homogeneous communities is slower, resulting in the formation of a large number of small, unconnected communities. More importantly, the scattered spread of positive seeds prevents mergers between the anti-vaccine communities. As a consequence, these smaller communities yield a smaller epidemic size. 

In contrast, targeted campaigns generate limited seeds as they work with a specific and static target set and not the entire population, so their impact is restricted to the positions of these seeds. In the centrality-based campaign, these positions are the most central nodes in the network, and as a result, they behave better in reducing the epidemic as they efficiently mitigate  the connectivity between anti-vaccine communities compared to the targeted random campaign, where the target set is chosen at random.

On the other hand, when the social rate is high, targeted campaigns work better in containing the disease dynamics, resulting in smaller epidemic size as the strategically positioned targets reduce the connectivity between the anti-vaccine communities.

To further explain, consider the scenario where $\mu^- = \mu^+$, represented by purple curves. In the random campaign shown in Fig  \ref{Fig3}A, the epidemic size increases as the social rate $\omega$ increases. However, the targeted random and targeted central campaigns shown in Fig \ref{Fig4}A and Fig \ref{Fig4}C, respectively, exhibit an inverse behavior, with the epidemic size decreasing as the social rate increases. The reason for this is that with broad seeding and less social interaction, the seeds act as obstacles distributed across the network, efficiently mitigating the merging between anti-vaccine communities. However, with a limited number of targets in the targeted campaign and a low rate of social interaction, these campaigns fail to mitigate the propagation of anti-vaccine influence, which is reinforced by broad and continuous exposures, as their effect only associates with their positions. On the other hand, with higher social rates, opinions are diffused faster, making it challenging to exert more influence over time. Thus, the random campaign becomes less efficient compared to the targeted campaign, where the fixed positions of the targets successfully impede the connectivity of the anti-vaccine communities. The same behavior is observed  regardless of the size of the positive campaign budget $\mu^+$.

Furthermore, Fig \ref{Fig4}B and Fig \ref{Fig4}D display the system at time $\tau=400$, allowing for a comparison between the campaign and the anti-vaccine opinion only scenario for the targeted random and centrality-based campaigns, respectively. The figure distinctly illustrates that the propagation of positive vaccine-related information results in a positive effect. As demonstrated in the figures, this approach reduces the epidemic size when compared to the scenario where only anti-vaccine opinions exist. Furthermore, the suppression of the epidemic increases as the positive general exposure rate increases. However, as the social rate increases, the growth of communities also increases, leading to a higher epidemic size. 

A noteworthy observation is that the centrality-based campaign is more effective in reducing the epidemic size than random campaigns compared to the scenario where only anti-vaccine opinions are spread. Additionally, the targeted random campaign is  more effective than the random campaign. For instance, consider the scenario where $\mu^+ = \mu^-$, as shown by the purple lines in Fig \ref{Fig3}B, Fig \ref{Fig4}B, and Fig \ref{Fig4}D. At the highest social rate, i.e., $\omega= 10^{-2}$, the epidemic size for the random, targeted random, and centrality-based campaigns is 1615 $\pm$ 27, 288 $\pm$ 9, and 161 $\pm$ 5, respectively, compared to 4174 $\pm$ 15 for the scenario where only anti-vaccine opinions are being spread.

\subsection*{Results for the dynamic campaigns}
This section presents the epidemic size obtained by applying targeted dynamic campaigns to disseminate positive vaccine-related information. This includes the dynamic random and dynamic local information-based (Locl-info) campaigns.  

\begin{figure}[!h]
% \centering
	
	\centering

    \includegraphics[width=\linewidth]{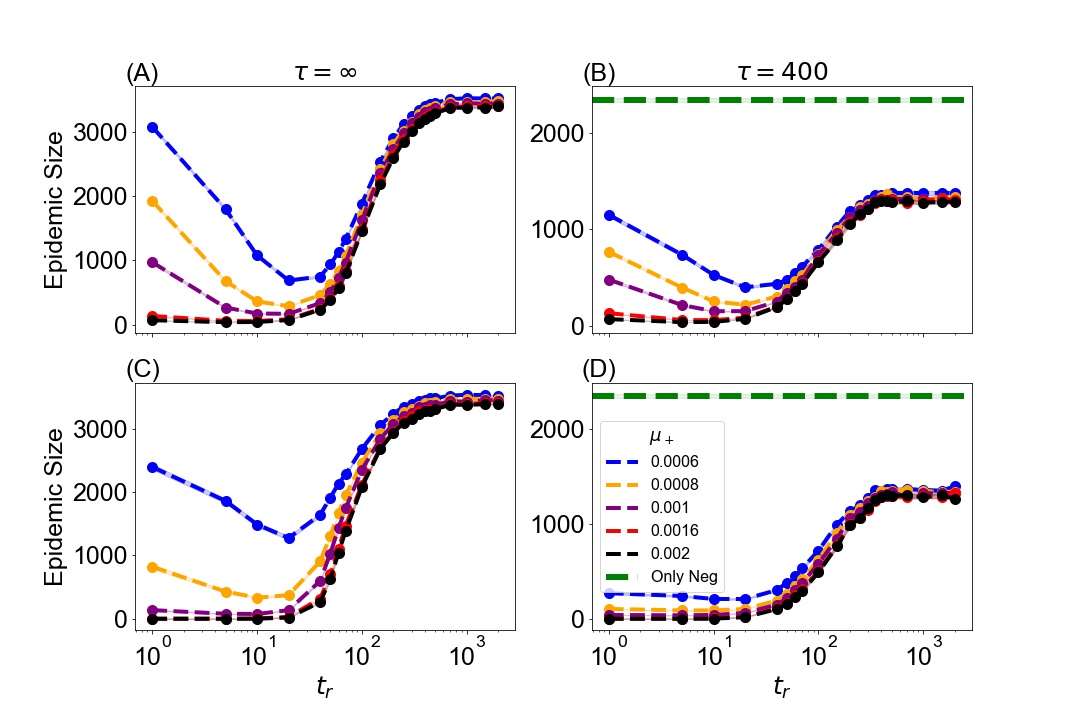}
    
\captionsetup{justification=justified, singlelinecheck=false}
\caption{{\bf Dependence of the average epidemic size on the campaign updating interval using dynamic campaigns.}(A) and (B) represent the dynamic random campaign, and (C) and (D) represent the Locl-Info campaign. The left panels illustrate the long-run scenarios, while the right panels illustrate the short-run scenarios.  The targets set $T=50$,  social rate is $\omega^+=\omega^-= 0.006$. The figures show different
positive exposure rates $\mu^+$ with fixed negative exposure rate $\mu^-=0.001$.}
\label{Fig5}
\end{figure}

 Fig \ref{Fig5} displays the results obtained by these campaigns. We evaluate the campaigns for different update times and results are presented as a function of the target set update time interval $t_r$. Here if the target set is changed very often, since multiple exposures are required for adopting opinions, the likelihood of each individual agent being influenced is very low. Correspondingly, fairly small amounts of influence are spread over a large set of agents that are targeted at different times. In contrast, when leaving the target set unchanged for longer, agents in the target set can accumulate multiple exposures which might lead to opinion adoption. However, this also implies that influence is not spread very widely and occasionally agents who already hold an opinion might be targeted. From these considerations it becomes clear that there must be an optimal switching time which maximizes the effect of the positive campaign.

 Fig \ref{Fig5}A and Fig \ref{Fig5}C illustrate the long-run scenario with $\tau = \infty$ for varying update time intervals for the dynamic random and Locl-Info campaigns, respectively. The results demonstrate that an optimal update time interval exists at around $t_r=20$. This is particularly obvious when the general positive exposure rate $\mu^+$ is lower than the negative one $\mu^+ \ll \mu^-$.
 This optimal time results from the trade-off explained above. Therefore, with a lower general positive exposure rate, the campaign performs better with relatively slow updates. 

On the other hand, as the general positive exposure rate increases, indicating a stronger positive influence, the pronounced effect—i.e., an optimal time at $t_r = 20$—diminishes. This effect becomes almost negligible when $\mu^+ > \mu^-$ in the dynamic random campaign, and when $\mu^+ \geq \mu^-$ in the Locl-Info campaign.  In such cases, the fastest update strategy $t_r=1$ becomes an effective option. This phenomenon occurs because a higher general positive exposure rate increases the probability of positive influence, and when combined with quick updates, it allows us to attain widespread coverage by targeting susceptible agents before they are negatively influenced. Moreover, after a time interval of $t_r=10$, the epidemic size continues to increase as the interval $t_r$ increases until reaching a stationary state, where there are no further changes in the epidemic size. As we increase the update intervals, we reduce the scope of our targeting coverage, resulting in a corresponding decrease of the positive effects we initially achieved in mitigating the negative influence.

Furthermore, comparing the dynamic random and Locl-Info campaigns, Fig \ref{Fig5}A and  Fig \ref{Fig5}C demonstrate varying performance in reducing the spread of an epidemic. For the smallest positive general exposure rates displayed in the figure, e.g., blue lines and $\mu^+=0.0006$, which is much smaller than the negative general exposure rate $\mu^- =0.001$, the dynamic random strategy outperforms the Locl-Info strategy at the optimal time $t_r=20$, resulting in a smaller epidemic size. At time $t_r=20$, the epidemic size is $1276$ $\pm$ $29$ and $688$ $\pm$ $19$ with $\mu^+ = 0.0006$, and $375$ $\pm$ $13$ and $288$ $\pm$ $8$ with $\mu^+ = 0.0008$ for Locl-Info and dynamic random, respectively. However, as the general positive exposure rate increases, the Locl-Info strategy is more effective at mitigating the negative influence. This is also noticeable when both positive and negative rates spread at the same rate, i.e., $\mu^+ = 0.001$, where epidemic size at time interval $t_r = 1$, is $139$ $\pm$ $8$ and $972$ $\pm$ $23$ for Locl-info and dynamic random, respectively. Furthermore, the epidemic size reduced even further when the positive general exposure rate was much greater than the negative rate with $\mu^+ =0.002$ and $\mu^-=0.001$ and reached $3$ $\pm$ $0.08$ and $70$ $\pm$ $2$ for the Locl-info and dynamic random campaigns respectively, at $t_r=1$.

 Fig \ref{Fig5}B and  Fig \ref{Fig5}D display the system at time $\tau=400$ for the dynamic random and Locl-Info campaigns, respectively, allowing for a comparison between the campaigns and the only anti-vaccine opinion scenario. Once again, the dissemination of positive vaccine-related information yields a positive impact in mitigating the diffusion of anti-vaccine sentiments compared to the scenario where only anti-vaccine opinions are disseminated. Furthermore, the suppression of the epidemic increases as the positive general exposure rate increases.

\subsection*{Results for the advanced local-info campaigns}
\begin{figure}[!h]

	\centering

	\includegraphics[width=\linewidth]{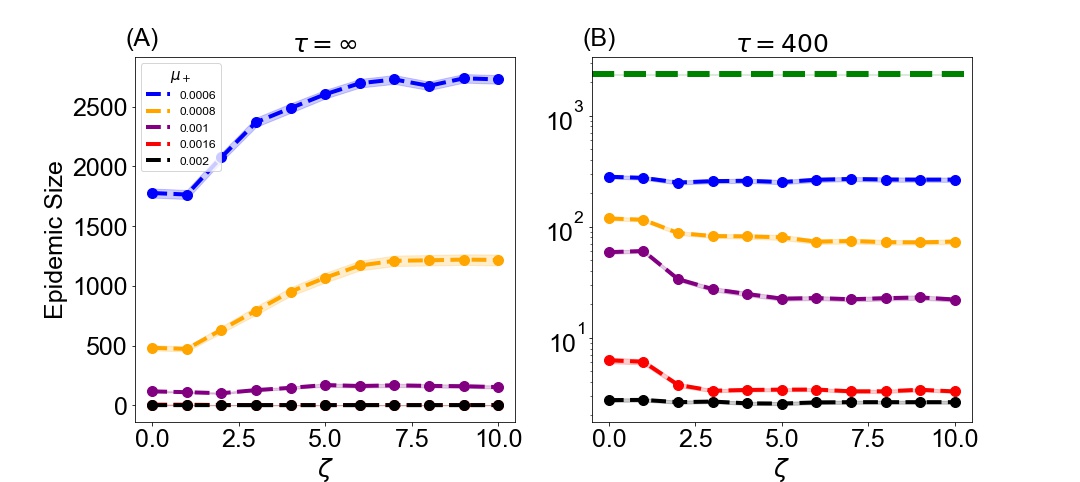}

\captionsetup{justification=justified, singlelinecheck=false}
\caption{{\bf Dependence of the average epidemic size on the number of anti-vaccine neighbors $\zeta$ using the advLocl-Info campaign.} The updating time is $t_r = 1$, the social rate is $\omega = 0.006$  for both negative and positive $\omega^+ = \omega^-$, and the size of the target set is $T=50$. The figures show different positive exposure rates $\mu^+$ with fixed negative exposure rate $\mu^+ =0.001$.}
\label{Fig6}
\end{figure}

\begin{figure}[!h]
% \centering  

	\centering         
	\includegraphics[width=\linewidth]{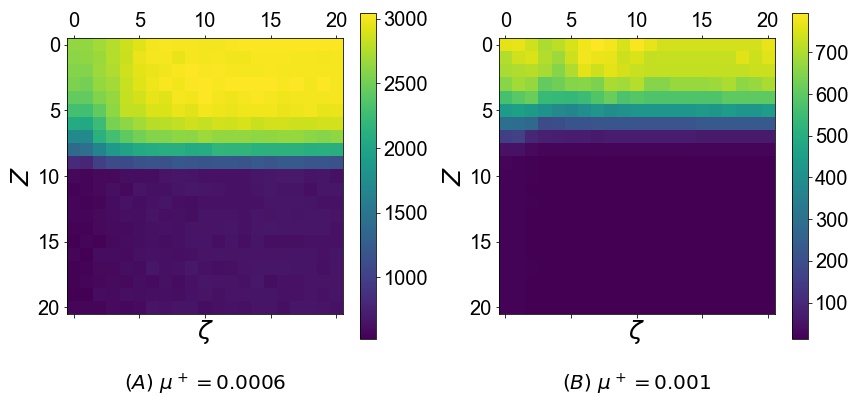}

\captionsetup{justification=justified, singlelinecheck=false}
\caption{{\bf Average epidemic size using the multi-objective advanced advMultLocl-Info  dynamic campaigns in the long-run setting of $\tau = \infty$.} The figure shows the performance of the dynamic campaigns with $T = 50$ targets. The epidemic size is shown as a function of the number of negative neighbors $\zeta$ and the number of neutral neighbors $Z$ a neutral has at time $t$. The updating time is $t_r$ =1, and the social rate is $\omega = 0.006$ for both negative and positive $\omega^+=\omega^-$. The general exposure influence rate for negative is $\mu^- = 0.001$ and the positive rate is shown in the figures captions.}
\label{Fig7}
\end{figure}

This section presents the epidemic size obtained when applying the advanced targeted dynamic campaigns to disseminate positive vaccine-related information. This includes the targeted advanced local-info (advLocl-Info) and  multi-objective advanced Locl-Info (advMultLocl-Info) campaigns.  For the advLocl-Info campaign, the results are presented as a function of the target number of anti-vaccine neighbors $\zeta$ to assess the impact of this parameter on mitigating the growth of anti-vaccine communities and, consequently, reducing the size of the epidemic. For the advMultLocl-Info campaign, the results are presented  as a function of both $\zeta$ and $Z$,  with different values of $\mu^+$ to explore the trade-off between maximizing the growth of pro-vaccine communities and minimizing the growth of anti-vaccine communities. The results are shown in Fig \ref{Fig6} and Fig \ref{Fig7} for advLocl-Info and  advMultLocl-Info campaigns, respectively.  

For advLocl-Info campaign, considering the long-run behavior with $\tau=\infty$, see Fig \ref{Fig6}A, we found that optimal performance is obtained for $\zeta \leq 1$, which  is particularly obvious when the positive rates $\mu^+$ are significantly lower than the negative rates $\mu^-$, as indicated by the blue line in the graph. We also observed the same behavior found in  the dynamic campaigns, where fast updates with weak influence rates are not effective in convincing the target set before they are updated since agents require multiple exposures to adopt a particular opinion. The challenge becomes greater when selecting neutral agents who are surrounded by many anti-vaccine adopters, as they are more likely to be negatively influenced by their social contacts. However, this behavior diminishes as the positive influence rate increases and leads to a significant reduction in the epidemic size, particularly when $\mu^+ \gg \mu^-$, regardless of the $\zeta$ value. For example,  the epidemic size remains the same at $\zeta = 1$ and $\zeta = 8$, taking a value of 3 $\pm$ 0.1, see Fig \ref{Fig6}A at $\mu^+$ = 0.002.

In the case of the advMultLocl-Info campaign, similar to advLocl-Info, we have observed that an optimal reduction in the epidemic size is obtained for $\zeta \leq 1$, particularly when the positive rates $\mu^+$ are considerably lower than the negative rates $\mu^-$, as depicted in Fig \ref{Fig7}A. Moreover, this reduction increases as $Z$ increases, indicating a focus on targeting individuals with a larger number of neutral neighbors. However, there is a remarkable  observation that when $Z\geq10$, the epidemic size is significantly diminished irrespective of the $\zeta$ value, which is equivalent to the average degree of the network $\langle k \rangle=10$. Furthermore, when $Z>10$, the epidemic size remains relatively constant.

Furthermore, in the case of $\mu^+ = \mu^-$, as shown in Fig \ref{Fig7}B, a more significant reduction in the epidemic size is observed with fewer target number of  neutral neighbours $Z$ compared to the previous scenario. The figure indicates that we can achieve a greater reduction when $Z>7$. As an example, this campaign efficiently reduced the epidemic size to  12 $\pm$ 0.5 when $\zeta=Z=10$. Furthermore, beyond this point, this reduction remains relatively constant, and increasing the $Z$ value does not yield any additional effect. In this scenario, the number of anti-vaccine neighbors, does not have a major effect.

\begin{figure}[!h]
% \centering  

	\centering         
	\includegraphics[width=\linewidth]{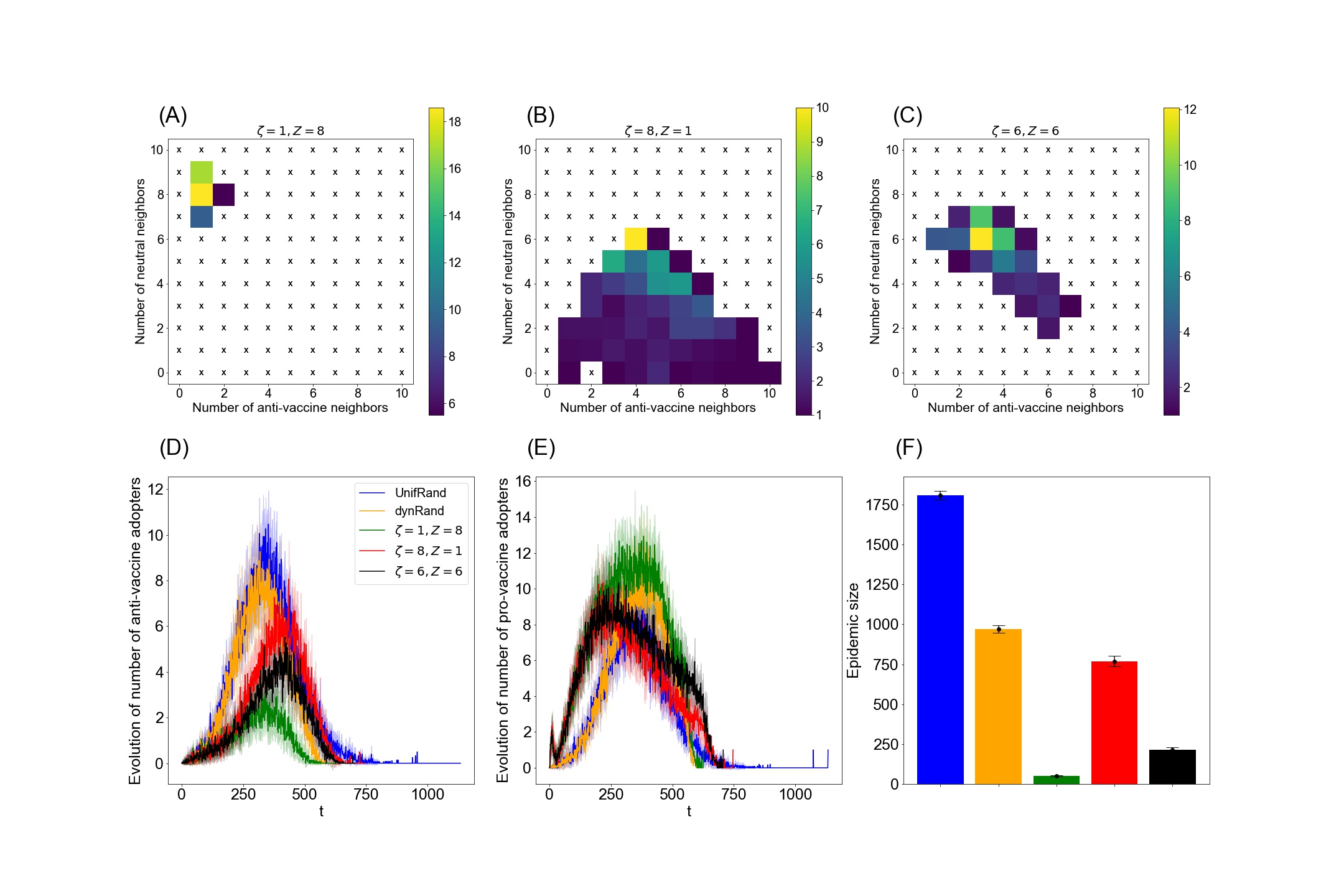}

\captionsetup{justification=justified, singlelinecheck=false}
\caption{{\bf Targeting scheme for  advMultLocl-Info campaign.}  The top panels illustrate the neighborhood structure of the target set at a single time step t=350 during the opinion diffusion stage, specifically showing the number of anti-vaccine neighbors and pro-vaccine neighbors. These panels represent the average number of agents with x anti-vaccine neighbors and y neutral neighbors for various settings: (A) $\zeta=1, Z=8$, (B) $\zeta=8, Z=1$, (C) $\zeta=6, Z=6$. (D) and (E) illustrate the evolution of anti-vaccine opinion adopters and pro-vaccine opinion adopters, respectively, while panel  (F) represents the corresponding epidemic size.  `x' indicates that no agent exists for that neighborhood pattern. The analysis is the average of 15 simulations.}
\label{Fig8}
\end{figure}

To elucidate the targeting strategy of this campaign, we illustrate in Fig \ref{Fig8} the neighborhood structure for the targeted agents of this campaign in conjunction with the evolution of opinion diffusion for both anti- and pro-vaccine adopters. In this campaign, by choosing a higher target number of neutral neighbors $Z$ than the target number of anti-vaccine neighbors $\zeta$, we prioritize the number of  neutral neighbors over the number of  anti-vaccine neighbors and vice versa. In the former scenario, this strategic shift directs positive allocation to agents more likely to propagate positive influence to a greater extent, as they have a larger number of neutral neighbors than anti-vaccine neighbors. This potentially will maximize the growth of pro-vaccine communities while restricting the growth of the anti-vaccine communities by being in proximity to them. In Fig \ref{Fig8}A, an illustrative example of such a scenario is presented, with $Z=8$ and $\zeta=1$. As observed in the figure, this scenario focuses on targeting agents with the highest number of neutral neighbors and the smallest number of anti-vaccine neighbors. The corresponding evolution of opinions, depicted in Fig \ref{Fig8}D and E, demonstrates that this targeting scheme maximizes the number of pro-vaccine adopters while simultaneously minimizing the number of anti-vaccine neighbors, as indicated by the green lines. Consequently, it leads to a greater reduction in the epidemic size, as depicted in Fig \ref{Fig8}F with the green bar.

Conversely, when the target number $Z$ is smaller than $\zeta$, we prioritize agents with a higher number of anti-vaccine neighbors than neutral neighbors. In this instance, we shift the targeting to focus on agents with a higher number of anti-vaccine neighbors and a relatively smaller number of neutral neighbors, see Fig \ref{Fig8}B for an example with $Z=1$ and $\zeta=8$. Although emphasizing agents with more anti-vaccine neighbors can protect them from the anti-vaccine influence, it is less efficient in mitigating the overall propagation of this influence in the network.  As depicted in Fig \ref{Fig8}D and E, this targeting scheme results in a higher number of anti-vaccine adopters and a lower number of pro-vaccine adopters compared to the first scenario (i.e., $Z=1$ and $\zeta=8$), as seen in the green and red lines. Correspondingly, it leads to a higher epidemic size, even though we are protecting the most vulnerable agents. 

Moreover, in Fig \ref{Fig8}C, we depict a scenario that assigns high priority to both $Z$ and $\zeta$, where $Z=\zeta=6$. In this scenario, as observed, we target agents with a relatively high number of neutral neighbors and a high number of anti-vaccine neighbors, avoiding the smaller numbers of neutral neighbors targeted by the scenario $Z=1, \zeta=8$. Consequently, this approach proves more efficient than the scenario depicted in Fig \ref{Fig8}B, resulting in lower numbers of anti-vaccine adopters, as depicted in the black line in Fig \ref{Fig8}D, and accordingly, a smaller epidemic size in Fig \ref{Fig8}F with the black bar. Despite this improvement over the previous scenario, the first scenario prioritizing neutral neighbors (Fig \ref{Fig8}A)  yields the best mitigation of negative influence and reduction in epidemic size. We further compare all three scenarios to random and dynamic random campaigns, depicted by the blue and orange colors, respectively, in opinion evolution and the corresponding epidemic size. As demonstrated in the figure, the advMultLocl-Info scenario outperforms  the benchmark cases in all scenarios.

\subsection*{The impact of the target set size}

\begin{figure*}[h]
\centering
        
\includegraphics[width=\linewidth]{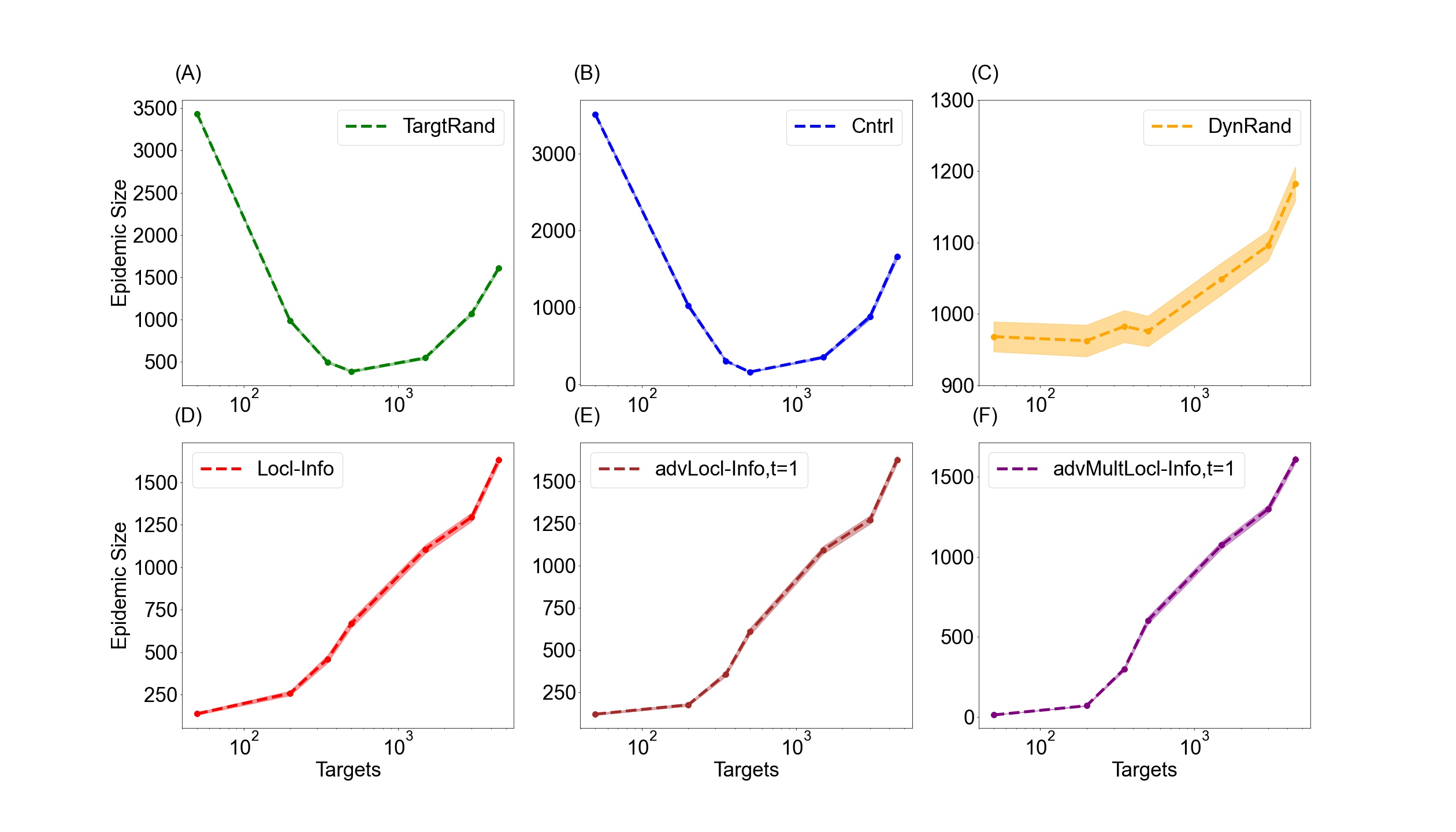}

\caption{{\bf Epidemic size obtained with varying target sizes for targeted campaigns in the long-run setting where $\tau=\infty$.} The figure illustrates the  epidemic size obtained for all campaigns with general exposure rates $\mu^- = \mu_+ =0.001$. The social rate is $\omega^+=\omega^+=0.006$ for all scenarios.  For dynamic campaigns, the updating time interval is $t$ =1, for  advLocl-Info $\zeta =1$, and advMultLocl-Info campaigns the target numbers of negative and neutral neighbours are $\zeta =10, Z =10$. For each scenario we generate 500 different networks, and perform 500 SIR model runs for each network.}
\label{Fig9}
\end{figure*}

The size of the target set is a crucial factor in determining the effectiveness of the targeted campaigns. Fig \ref{Fig9} demonstrates the correlation between the size of the target set and the extent of the epidemic in targeted campaigns.  In the case of static campaigns, see Fig \ref{Fig9}A and Fig \ref{Fig9}B, a larger target set results in better mitigation of anti-vaccine diffusion than a small target set. This is due to the positive allocation sticking around only these agents, and a relatively larger set ensuring a fair coverage in the network. However, we observed that a very large target set might act as noise and impede the focus of the campaign. For example, in the centrality-based campaign, after a certain point, i.e., T=500 which represents $10\%$ of the population, the curve starts to increase again due to the inclusion of numerous low centrality agents in the target set, which hampers the centrality effects and introduces more randomness in the selection process.

In dynamic campaigns, on the other hand, the results have shown that smaller target group sizes, particularly in Locl-Info, advLocl-Info, and advMultLocl-Info, yield better reduction compared to larger sizes, as depicted in Fig \ref{Fig9}(D-F). The dynamic random campaign demonstrates a slight increase as the target size increases. This suggests that it is more effective to direct resources to a smaller yet changeable target set. This observation can be attributed to the fact that a smaller size allows for more selective targeting of individuals who meet the campaign's criteria and are more likely to be influenced. On the contrary, a larger target set may include individuals who are less susceptible to the negative influence, introducing more randomness in the selection process and  resulting in decreased effectiveness. Furthermore, the positive strength allocated to a campaign is typically distributed evenly across the target group. As a result, a change in the target size can lead to a change in the strength allocated per individual. 

\subsection*{Discussion} \label{disscu-sec}

In this paper, we have investigated the impact of different strategic positive campaigns for spreading positive vaccine-related information to combat the spread of negative vaccine-related information. We also examined how these campaigns affect the diffusion of anti-vaccine opinions and the connectivity between  emerging anti-vaccine communities, thereby leading to a reduction in the epidemic size. We demonstrated that the existence of positive influence propagation has a positive impact on mitigating the flow of negative influence, leading to improved vaccination coverage and,  as a result, reduces the epidemic size. However, this impact varies across different campaigning approaches. 

One crucial factor in the diffusion of anti-vaccine sentiments within a social network is the level of social influence between individuals; as the social contagion rate increases, the size of the anti-vaccine communities grows, consequently increasing the epidemic size. This phenomenon can be observed in Figure \ref{Fig3}, specifically within the benchmark case where the scenario involves only anti-vaccine diffusion (depicted by the green line), and it persists despite the concurrent presence of pro-vaccine propagation through a random campaign. These results are consistent with previous studies that investigated a similar problem and demonstrated the role of social interactions in promoting the growth of anti-vaccine communities, thus increasing the epidemic size \cite{campbell2013complex,dorso2017vaccination}. We then demonstrated that targeted campaigns can effectively contain the spread of anti-vaccine diffusion in such a scenario (see Fig \ref{Fig4}).

\begin{figure}[!h]
\centering

\centering
    \includegraphics[width=\linewidth]{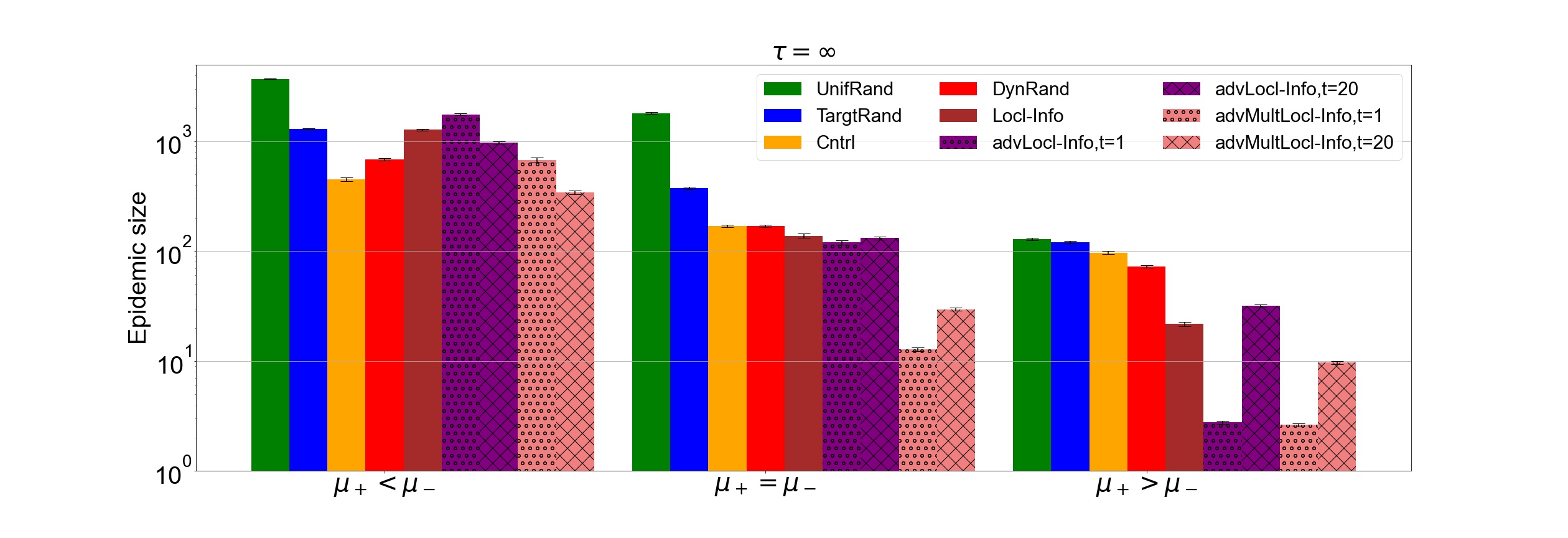}

\captionsetup{justification=justified, singlelinecheck=false}
\caption{{\bf Cross-campaign epidemic size comparison with $\tau=\infty$.} The figure illustrates the  epidemic size obtained for all campaigns with different positive rates $\mu^+$ compared to negative rate $\mu^-$. For all campaigns,  $\mu_- =0.001$. First group: $\mu_+ =0.0006$, second group: $\mu_+ =0.001$, and third group: $\mu_+ =0.002$. Social rate is $\omega=0.006$ for all scenarios. Target set size $T=500$ for static campaigns, i.e., TargtRand, and Cntrl, and $T=50$ for the other dynamic campaigns.  For dynamic campaigns, the updating time is $t_r = 20$, in addition we include $t_r=1$ for  advLocl-Info and advMultLocl-Info campaigns. In addition, for advLocl-Info $\zeta$ =1,  for advMultiLocl-Info $\zeta$ = $Z$ =10.}
\label{Fig10}
\end{figure}

In Fig \ref{Fig10}, we compare the best scenarios for each campaign, depicting three distinct states of strength allocation in the general exposure for negative and positive campaigns, considering a high social rate—a crucial scenario in which anti-vaccine communities can expand significantly. The aim is to investigate which campaign can mitigate this expansion most effectively. The figure also demonstrates the long-run behavior of the system. Across all scenarios, advanced local information campaigns, particularly advMultiLocl-Info, achieve the best performance in reducing the epidemic size.

In situations where the positive general exposure rate is much smaller than the negative general exposure rate (i.e., $\mu^+ \ll \mu^-$), the best practice is to use advMultiLocl-Info with slower updates (i.e., $t_r=20$) if we assume complete knowledge of the population's vaccine-related attitudes. The next best option is the centrality-based campaign when we have incomplete knowledge about vaccine attitudes. The former results in an epidemic size of $345$ $\pm$ $13$, while the latter results in $449$ $\pm$ $17$.

When both negative and positive general exposures exert the same rate of influence (i.e., $\mu^+ = \mu^-$), advMultiLocl-Info with fast updates (i.e., $t_r=1$) is the most effective, while centrality-based and other dynamic campaigns produce relatively similar performance. Finally, when the positive general exposure rate is higher than the negative one (i.e., $\mu^+ > \mu^-$), the best practice is to use either advLocl-Info or advMultiLocl-Info with fast updates (i.e., $t_r=1$).

From Fig \ref{Fig10}, it is observed that although the random strategy (UnifRand) shown by green bars reduced the epidemic size, it is not the most efficient strategy compared to other campaigns, as the resulting epidemic size is higher than that of the others. The static random campaign (TargRand) shown by blue bars follows a similar pattern but performs better than the random campaign. This observation, as investigated earlier, is due to the fact that the random campaign (UnifRand) fails to exert more influence over time, and opinions become socially-driven. In contrast, the targeted random campaign (TargRand) targets fixed positions in the network, creating barriers that prevent the clustering of anti-vaccine communities, making it slightly more effective than the random campaign (UnifRand).  This observation is consistent with the study conducted by \cite{zhang2015limiting}, where the authors demonstrated that a random selection of the seed set resulted in poor mitigation of misinformation propagation.

A targeted campaign based on the network structure, specifically aimed at individuals with high centrality, proves effective in reducing the spread of negative opinions and mitigating social contagion. This approach places positive seeds strategically on the most central bridges within the network, thus preventing the merging of anti-vaccine communities more efficiently. However, this method requires a large number of targets to be efficient, making it impractical in limited-resource settings, see Fig \ref{Fig8}B. 

As demonstrated above selecting targets based on local negative information proves to be the most effective strategy. By focusing on individuals who are more vulnerable to negative influences, this technique outperforms random selection and even outperforms the centralized approach. It also demonstrates efficacy even with a small target set. Quantifying the amount of negative influence on neutral social contacts yields further reductions in the size of  the epidemic. Moreover, incorporating the potential for maximizing positive influence, by considering the number of neutrals for each candidate, leads to higher effectiveness, as seen in  advMultiLocl-Info strategy. However, it is important to note that these local information-based campaigns assume complete knowledge of the vaccine-related attitudes within the population.

Moving beyond the exploration of effective targeting strategies, several historical instances exemplify the detrimental effects of vaccine misinformation, where negative  beliefs have profoundly impeded public health efforts. The spread of misinformation during the COVID-19 pandemic resulted in increased vaccine hesitancy and low vaccination rate\cite{lee2022misinformation}. Similarly, during the Ebola outbreak in North Kivu, a notable association was observed between the mistrust in health systems and the belief in misinformation, which led to reduced inclination in adopting preventive behaviors, including acceptance of Ebola vaccines \cite{vinck2019institutional}.  Additionally, the year 2019, witnessed a resurgence of measles due to a significant decline in vaccine coverage, resulting from the spread of anti-vaccine attitudes as a significant cause  \cite{hotez2020combating}. Such cases exemplify the pivotal role of accurate information in managing and mitigating the effects of misinformation associated with various infectious diseases.

Locally tailored approaches to improving health promotion are vital. Our approaches here need further development, but the modelling is designed to be flexible to support local needs and thus can incorporate localised scenarios with regard to the rate of dissemination of good and bad public health information, and the number of positive, neutral and negative nodes. It can also be adapted to mimic online or offline dissemination.  In addition, the current model operates on the assumption of comprehensive knowledge regarding vaccine-related attitudes within the social network. Full knowledge of the attitude of every individual is not realistic in practice; however, obtaining some information is feasible from processes such as social network analysis of data from social media platforms. For instance, empirical studies leveraging data from online social networks have applied sentiment analysis and machine learning algorithms to categorize individuals’ attitudes into distinct states such as pro-vaccine, anti-vaccine, and neutral towards vaccination \cite{salathe2011assessing,d2019monitoring,abd2020sentiment,yousefinaghani2021analysis}. Additionally, public attitudes toward vaccines have been explored through the analysis of real-time, spatial-temporal, and socio-demographic data from social media, unveiling the spatial distribution of these attitudes \cite{hu2021revealing,umair2023sentimental,cheng2023exploring}. Such platforms and techniques facilitate real-time, socio-geographic monitoring of public attitudes, aiding health campaigns in implementing optimal interventions. 

The proposed heuristics can be further extended to a broader spectrum of opinion diffusion models. The present results are restricted to the assumption of static opinions. Nonetheless, we believe that these strategies could be effectively integrated within models permitting opinions to be switched back and forth, such as the voter model and epidemic models, e.g. SIS model. An additional limitation of this study is the singular network employed for both information and disease diffusion. Future research would benefit from exploring the implications of employing distinct network structures tailored to each diffusion process.

\section*{Conclusion}
The purpose of this study is to investigate effective mechanisms to mitigate the spread of anti-vaccine attitudes and reduce the size of epidemics by applying positive counter-campaigns that spread positive vaccine-related sentiments.  We proposed efficient heuristics to combat negative influence propagation. We have demonstrated that these campaigns can impede the flow of anti-vaccine attitudes and changed the distribution of unvaccinated individuals within the population, which in turn changed the structures of anti-vaccine communities and suppressed the spread of epidemics. Our study has proposed strategies based on two main paradigms: social network structure and negative local information, in addition to two control schemes. Through extensive experiments, we systematically studied and analyzed the performance of the proposed strategies in reducing the size of epidemic, identifying their strengths and limitations.

We have demonstrated that targeted campaigns that select a subset of the population based on certain criteria have been found to be more effective in suppressing the epidemic compared to random campaign, particularly in scenarios with high levels of social interactions. The latter approach results in poor mitigation compared to other methods. In contrast, centrality-based and local information-based strategies have shown superior performance. The centrality-based campaign targets central nodes which effectively hinder the merging of emerging communities, while the local information-based methods prevent the most vulnerable agents from being negatively influenced. Moreover, the dynamic control approach, which involves continuous updating of the target set, has been found to be more effective in suppressing the epidemic compared to a static control approach. This approach provides continuous and iterative exposure to positive messaging while keeping the campaign involved with the evolution of anti-vaccine attitudes. Furthermore, prioritizing those susceptible based on their neighborhood state performs even better in mitigating negative influence propagation.

\section*{Acknowledgments}
The authors acknowledge the use of the IRIDIS High Performance Computing Facility in the completion of this work.
\nolinenumbers

% Either type in your references using
% \begin{thebibliography}{}
% \bibitem{}
% Text
% \end{thebibliography}
%
% or
%
% Compile your BiBTeX database using our plos2015.bst
% style file and paste the contents of your .bbl file
% here. See http://journals.plos.org/plosone/s/latex for 
% step-by-step instructions.
% 

\end{document}